\documentclass[12pt,a4paper]{article}
\usepackage{amssymb,amsmath,amscd,epsfig}
\title{Neutrino magnetic moment signatures in the supernova neutrino signal}
\author{
 Oleg Lychkovskiy 
\thanks{e-mail: lychkovskiy@itep.ru}
\hspace*{2mm}$^{\rm a,b}$
\\ ${\rm ^a}$ {\small\it Institute for Theoretical and Experimental Physics}\\
{\small\it 117218, B.Cheremushkinskaya 25,
Moscow, Russia}\\
${\rm ^b}$ {\small\it Moscow Institute of Physics and Technology
}\\{\small\it 141700, 9, Institutskii per., Dolgoprudny, Moscow
Region, Russia}}

\begin{document}

\newcommand{\nue}{\nu_e}
\newcommand{\nul}{\nu_l}
\newcommand{\antinue}{\bar{\nu}_e}
\newcommand{\ar}{\rightarrow}
\newcommand{\C}{{\rm C}}
\newcommand{\Hy}{{\rm H}}
\newcommand{\F}{{\rm F}}
\newcommand{\Co}{{\rm Co}}
\newcommand{\N}{{\rm N}}
\newcommand{\Ox}{{\rm O}}
\newcommand{\Fe}{{\rm Fe }}
\newcommand{\be}{\begin{equation}}
\newcommand{\ee}{\end{equation}}
\newcommand{\bc}{\begin{cases}}
\newcommand{\ec}{\end{cases}}
\newcommand{\B}{{\bf B}}
\newcommand{\Bp}{{\bf B_\perp}}
\newcommand{\rb}{\right)}
\newcommand{\lb}{\left(}
\newcommand{\kpc}{{\rm kpc}}

\maketitle

\begin{abstract}
It is known that if neutrino is a Dirac fermion with magnetic moment, then $\nu_L\ar\nu_R\ar\nu_L$ transition of supernova neutrinos may occur. The first stage of such transition is due to the neutrino spin flip inside the hot dense supernova core, while the second one -- due to the neutrino spin precession in the galactic magnetic field on the way from the supernova to terrestrial detectors. This can result in the detection of 60-200 MeV neutrinos simultaneously with the "normal" supernova neutrino signal, which would be a smoking gun for the Dirac neutrino magnetic moment, $\mu$. We argue that in case of a nearby supernova explosion ($\sim$ 10 kpc away from the Earth) one may observe such high-energy events in Super-Kamiokande if $\mu \gtrsim 10^{-13} \mu_B,$ and in a Mt-scale detector if  $\mu \gtrsim 0.5 \cdot 10^{-13} \mu_B.$ Such an observation by itself, however, may be not sufficient to determine the value of the magnetic moment, because of the ignorance of the interstellar magnetic field. We point out that if in addition a deficit of the neutronization burst neutrinos is established, it would be possible to extract the value of the magnetic moment from observations. We also briefly discuss a possible Majorana magnetic moment signature due to $\nu_e\ar\bar{\nu}_{\mu,\tau}$ flip inside the supernova core.
\end{abstract}


\newline

 \section{Introduction}

It is natural for neutrinos to have magnetic moments. 
In the Standard Model (with massive Dirac neutrinos included) magnetic moment of the neutrino mass eigenstate $\nu_i$  is proportional to its mass $m_i$ and reads (see \cite{Lee:1977tib},\cite{Fujikawa:1980yx}) 

\be \label{SM magnetic moment}
\mu_i=\frac{3eG_{\rm F}m_i}{8 \sqrt{2} \pi^2 }=3.2 \cdot 10^{-19}\frac{m_i}{1~{\rm eV}} \mu_B.
\ee

In the extensions of the Standard Model neutrino magnetic moments may be orders of magnitude larger (see 
\cite{Vysotsky:2002yu} and references therein). The best direct experimental limit currently reads \cite{Beda:2007hf}
\be
\mu_{\nu_e} < 6\cdot 10^{-11} \mu_B,~~~~~~~~90\%{\rm CL.}
\ee

Astrophysical limits on the neutrino magnetic moment are two orders of magnitude better, though model dependent and less definite. Most of such limits come from considering energy losses due to emission of neutrinos for various types of stars, such as red giants and white dwarfs, as well as for collapsing supernovae. Such energy losses, being dependent on magnetic moment $\mu,$ should not be too high. 
This allows to restrict neutrino magnetic moments (see \cite{Vysotsky:2002yu} for the review and \cite{Kuznetsov:2007mp} for probably the most stringent limit of this type):
\be
\mu \lesssim (1 - 3) \cdot 10^{-12} \mu_B.
\ee

Explosion of a nearby supernova, SN1987A, provided  a possibility to constrain $\mu$ using non-observation of high-energy neutrinos ($E_\nu=O(100 {\rm MeV})$) in two water Cherenkov detectors, which were operational at that time, IMB and Kamiokande. Different aspects of this method were discussed and investigated in a number of papers \cite{Fujikawa:1980yx}, \cite{Dar:1987yv}-\cite{Notzold:1988kz}. 
The idea of the method is the following. Left electron neutrinos, created in the hot collapsing supernova core at the first second of the collapse, experience spin flips in collisions with charged particles, among which electrons and protons play the major role:
$$\nu_L +e \ar \nu_R +e$$
\be\label{spin flip reactions}
\nu_L +p \ar \nu_R+ p.
\ee
 Sterile right neutrinos escape from the core without energy loss.  Typical energies of this leaking neutrinos, $O(100 {\rm MeV}),$ are of order of typical energies of the electrons inside the core (compare with 10-40 MeV for "standard" supernova neutrinos). Then neutrinos travel through the galaxy to reach detectors at the Earth, and a backward transition, $\nu_R\ar\nu_L,$ may occur in the galactic magnetic field. If the neutrino magnetic moment is large enough, this would result in the detection of high-energy neutrinos\footnote{
In this letter "high energy neutrinos" mean 60--200 Mev neutrinos.}
  simultaneously with the "normal" supernova neutrino signal. As no such high-energy neutrinos were detected by Kamiokande and IMB, one can constrain neutrino magnetic moment \cite{Barbieri:1988nh}\cite{Notzold:1988kz}:
\be
\mu \lesssim  10^{-12} \mu_B.
\ee
Here it is supposed that one half of right neutrinos turned to left neutrinos in the galactic magnetic field.

In this letter we discuss prospects of exploring neutrino magnetic moment in the future collapse of a nearby Supernova. We concentrate mainly on the case of Dirac neutrinos, and only in section 3 briefly comment upon one possible Majorana neutrino signature. We estimate the expected number of high-energy neutrino events in Super-Kamiokande (22.5 kt fiducial mass) and in a future Megaton-scale detector. It is found that, for a conservative estimate, Super-Kamiokande may be sensitive to $\mu \gtrsim 10^{-13} \mu_B,$ while a 0.5 Mt water detector -- to  $\mu \gtrsim 0.5\cdot 10^{-13} \mu_B.$\footnote{
For 
$\mu \lesssim 10^{-13} \mu_B$ the number of events in the detector is proportional to $\mu^4$ (see section 2.1). This explains why increasing the fiducial mass by factor 20 results only in doubling of the sensitivity.} It is also pointed out that the deficit of neutronization burst neutrinos due to $\nu_L\ar\nu_R$ transition in the interstellar magnetic field may be another clear signature of Dirac neutrino magnetic moment. If both deficit of neutronization burst neutrinos and high energy neutrinos are observed, this may allow to determine the value of the neutrino magnetic moment.  


Two remarks should be made. First, in principle it is possible that the backward transition, $\nu_R\ar\nu_L,$ 
occurs already in the star envelope, if a strong magnetic field is present there. In this case high-energy reflipped left neutrinos interact with the envelope matter and release their energy. This helps to produce a successful supernova explosion, but cancels the high energy neutrino burst. It was Dar \cite{Dar:1987yv} who first pointed out on this  possibility. The effective matter potential, which is felt by left neutrinos and is not felt by right neutrinos, in general suppresses $\nu_R\ar\nu_L$ transition in the magnetic field inside the star; nevertheless it may vanish at some radius, thus allowing $\nu_R\ar\nu_L$ transition. This question was discussed in a number of papers \cite{Voloshin:1988xu}--\cite{Blinnikov:1988xq}. We suppose that $\nu_R\ar\nu_L$ transition {\it does not occur} inside the star. According to \cite{Voloshin:1988xu}, a sufficient condition for this is 
\be\label{Voloshin condition} 
\mu B \lesssim  (10^{-12} \mu_B)\cdot(10^{13}~ {\rm Gs}),
\ee
 $B$ being the magnetic field at $\sim 100$ km from the center of the star.   

The second remark concerns the exact meaning of $\mu.$ In general, electromagnetic properties of neutrinos are described by $3\times  3$ complex matrix $M_{em}$, which embraces diagonal (for Dirac neutrinos) and transition (both for Dirac and Majorana neutrinos) magnetic and electric dipole moments 
\cite{Voloshin et al}. For ultra-relativistic neutrinos the phenomenological manifestations of electric dipole moments coincide with those of magnetic dipole moments \cite{Okun:1986uf}. Bearing this in mind, we do not consider explicitly electric dipole moments.  Dealing with Dirac neutrinos, we assume for simplicity  that $M$ is diagonal in any basis, i.e. $M_{em}=\mu\cdot{\bf 1},$ where ${\bf 1}$ is a unit matrix and $\mu$ is a universal magnetic moment.

The paper is organized as follows. In section 2 we obtain our main results. We proceed in two steps: first we separately discuss neutrino propagation in the interstellar space (section 2.1), and then calculate the expected number of high-energy events in the detector (section 2.2). In section 3 we discuss the uncertainties involved and prospects for improving the accuracy of the results; also we make some remarks concerning diffuse neutrinos from all past supernovae, as well as possible Majorana neutrino magnetic moment signature. The results are summarized in section 4.

\section{Dirac neutrino magnetic moment signatures in the supernova signal}

\subsection{Neutrino propagation in the interstellar space and $\nu_R\ar\nu_L$ transition in the galactic magnetic field}

The interaction of the neutrino magnetic moment with the magnetic field $\B$ leads to the neutrino spin precession (or, in other words, to $\nu_R \leftrightarrow \nu_L$ oscillations), which is described (in the ultra-relativistic case) by\cite{Voloshin et al} 
\be \label{evolution eq} 
i\frac{d}{dx}\nu(x)= (E+\mu {\boldsymbol \sigma} {\Bp(x)} )\nu(x).
\ee
Here 
$ \nu (x)=\left(\begin{array}{l} \nu_L(x) \\ \nu_R(x)
\end{array}\right),$
$E$ is neutrino energy, $\boldsymbol\sigma$ is a vector constructed from Pauli matrices, and $\Bp(x)$ is a component of $\B$ normal to the neutrino momentum.
If for every $x$ magnetic field $\Bp(x)$ lies in the same plane, then the phase of oscillations is given by
\be\label{phase general}
\phi=\int  \mu B_\perp (x) dx.
\ee
The oscillation probability, $P(\nu_R\ar\nu_L)=\sin^2\phi,$ may be easily calculated for this case.  For the constant magnetic field
one gets
\be\label{phase}
\phi=\mu_\nu B_\perp x= 0.9 \lb \frac{\mu}{10^{-13}\mu_B} \rb \lb \frac{B_\perp }{{\rm \mu G}}\rb 
\lb \frac{x}{10~{\rm kpc}} \rb.
\ee

Galactic magnetic field has a complicated structure (see, for example, \cite{Vallee}, \cite{Han:2007uu}). Its typical strength is not less than 1 $\mu$G, and probably somewhat larger. It can be represented as the sum of regular (large-scale) and random (small-scale) components. Length scales of the random component are much smaller than 1 kpc, therefore, according to (\ref{phase}), this component is irrelevant for our purposes.\footnote{
Strictly speaking, relevance of the random magnetic field to the spin rotation is determined by 
$\gamma=\mu^2 \langle B^2 \rangle L_c x$ 
\cite{Nicolaidis:1991da},
where $L_c$ is a field length scale. Taking $\mu=10^{-12}\mu_B, ~x=10$ kpc and $B \sim 1~\mu$Gs, $L_c\sim 10 $ pc (see \cite{Han:2007uu} and reference therein), one obtais
$\gamma\sim 0.1\ll 1.$ This means that the effect of the random magnetic field on the spin precession may be neglected.
}
 Length scales of the regular component are of order of 1 kpc. In the galactic disk regular magnetic field is directed along the spiral arms, clockwise or counterclockwise depending on the spiral arm. There is a variety of galactic magnetic field models (see the above mentioned references). We use a model described in \cite{Vallee}. It reproduces the main qualitative features of the radial dependence of the magnetic field in the inner galaxy, i.e. two field reversals at $\sim 4.5$ and $\sim 6.5$ kpc from the center of the galaxy, as well as the characteristic strength of the magnetic field.

We consider a frequently discussed case of a supernova exploding in the inner part of the disk of our galaxy, $D=10$ kpc away from the Solar system. For simplicity, we assume that it is situated on the line which connects Solar system and galactic center.
The radial dependence of $B_\perp (r)$  reads \cite{Vallee}
\be \label{B regular}
B_\perp (r) = \bc 
0.9 ~\mu{\rm G} & 0~\kpc < r \le 2 ~{\rm kpc} \\
3.8 ~\mu{\rm G} & 2~\kpc < r \le 3~ {\rm kpc} \\
3.1 ~\mu{\rm G} & 3~\kpc < r \le 4~ {\rm kpc} \\
-2.2 ~\mu{\rm G} & 4~\kpc < r \le 5~ {\rm kpc} \\
-1.9 ~\mu{\rm G} & 5~\kpc < r \le 6~ {\rm kpc} \\
1.9 ~\mu{\rm G} & 6~\kpc < r \le 7~ {\rm kpc} \\
2.5 ~\mu{\rm G} & 7~\kpc < r \le 8~  {\rm kpc,} \\
              \ec
\ee 
Here $r$ is the galactocentric distance. The distance from the Sun to the galactic center is taken to be 7.2 kpc in \cite{Vallee}.

From (\ref{phase general}) and (\ref{B regular}) one obtains the probability of $\nu_R\ar\nu_L$ oscillations: 
\be\label{P}
P_{\nu_R\ar\nu_L}=\sin^2  \lb 1.1 \frac{\mu}{10^{-13}\mu_B}\rb. 
\ee

For $\mu \gtrsim   10^{-13}\mu_B$ the probability oscillates  rapidly with $\mu,$  phase being strongly dependent on the galactic magnetic field model. This means, in fact, that for such values of $\mu$ one should consider 
the phase $\phi$ as a uniformly distributed random value. In this case $P_{\nu_R\ar\nu_L}$ is also a random value. Its expectation value is 
$P_{av}=0.5$, and $P_{\nu_R\ar\nu_L}> 0.025 $ with $90\%$ probability.

One should take into account flavour transformations along with the spin precession in the problem involved. Right electron neutrinos, produced in a supernova, $\nu_{eR},$  quickly decohere into the the mixture of $\nu_{1R}$ and $\nu_{2R}$ (see, for example, \cite{Dighe:1999bi} for numerical estimates; some subtle details of this phenomenon were investigated in \cite{Dolgov:2005nb}\cite{Dolgov:2005vj}). The corresponding fractions in the mixture are equal to 
$\cos^2\theta_{12}$ and $\sin^2\theta_{12},$ $\theta_{12}\simeq 30^o$ being the mixing angle. In the interstellar medium  $\nu_{1R}\ar\nu_{1L}$ and $\nu_{2R}\ar\nu_{2L}$ transitions occur with the probability $P_{\nu_R\ar\nu_L}$ as described above. Finally, in the detector $\nu_{1L}$ and $\nu_{2L}$ show themselves as electron neutrinos with probabilities $\cos^2\theta_{12}$ and $\sin^2\theta_{12}$ correspondingly. Combining all the probabilities together, one finds the fraction $\kappa$ of right electron neutrinos, $\nu_{eR},$  which are converted to left electron neutrinos, $\nu_{eL},$ in the interstellar space on the way from the supernova to the Earth:
\be
\kappa =(\cos^4\theta_{12} + \sin^4\theta_{12})P_{\nu_R\ar\nu_L}=(1-0.5\sin^2 2\theta_{12})P_{\nu_R\ar\nu_L}
\approx 0.6 P_{\nu_R\ar\nu_L}.
\ee

\subsection{Dirac neutrino magnetic moment signatures in the detectors}

\begin{table}[t] 
\begin{center}
\begin{tabular}{|l|c|c|c|}
\hline
 & fiducial mass  & number of events,          & number of events,          \\
 &  of H$_2$O     & $\mu\gtrsim  10^{-13}\mu_B$ & $\mu\lesssim  10^{-13}\mu_B$ \\
\hline
Super-Kamiokande & 22.5 kt & $1.4~ P_{\nu_R\ar\nu_L} \lb\frac{\mu}{10^{-13}\mu_B}\rb^2$ & $\lesssim 1$   \\
\hline
MEMPHYS          & 440 kt  & $26 ~P_{\nu_R\ar\nu_L} \lb\frac{\mu}{10^{-13}\mu_B}\rb^2$  & 
$32 \lb\frac{\mu}{10^{-13}\mu_B}\rb^4$\\
\hline
\end{tabular}
\end{center}
\caption{Estimated numbers of events in Super-Kamiokande and MEMPHYS for the supernova in the inner part of the galaxy, 10 kpc away from the Solar system. In case $\mu\gtrsim 10^{-13}\mu_B$ the present knowledge of the galactic magnetic field does not allow to obtain any reliable value for the phase of spin precession 
$\phi;$ {\it a priory} the probability of 
$\nu_R\ar\nu_L$ conversion,  $P_{\nu_R\ar\nu_L}=\sin^2\phi,$ may take any value between 0 and 1.} 
\end{table}

In ref.\cite{Barbieri:1988nh} an expression for the number of events with energies greater than 60 MeV in the water Cherenkov detector is suggested\footnote{To be precise, our eq.(\ref{event rate}) may be obtained from eq.(11) in ref.\cite{Barbieri:1988nh}, if one introduces conversion probability $\kappa$ in their expression and conservatively chooses the $\nu_R$ emission time to be 0.5 s (it is considered to be (0.5-1)s in \cite{Barbieri:1988nh})}:
\be\label{event rate}
N=(0.06-6)\cdot P_{\nu_R\ar\nu_L}  \lb\frac{\mu}{10^{-13}\mu_B}\rb^2 \lb\frac{M_{\rm H_2O}}{1~{\rm kt}}\rb \lb\frac{D}{10 ~{\rm kpc}}
\rb^{-2}.
\ee 
The two-orders uncertainty mainly comes from the approximate treatment of state of matter in the hot supernova core. We postpone the discussion of this point to the next section, and conservatively use the lower value of $N.$  The estimated numbers of events for Super-Kamiokande and for the proposed $\sim 0.5$ Mt detector MEMPHYS\cite{Autiero:2007zj}\footnote{There are other Mt-scale water detector proposals, e.g. Hyper-Kamiokande\cite{Nakamura:2003hk} and UNO\cite{Wilkes:2005rg}. Planned sensitivities of all such detectors are comparable.}
 are presented in Table 1.  To obtain this estimate in case  $\mu\lesssim  10^{-13}\mu_B$ we substitute sine by its argument in eq.(\ref{P}).


As was mentioned in section 2.1, it is extremely hard to estimate the probability of $\nu_R\ar\nu_L$ transition, $P_{\nu_R\ar\nu_L},$ in case when $\mu\gtrsim  10^{-13}\mu_B.$   However, it is interesting to note that a supernova explosion may provide an independent possibility to {\it measure}  $P_{\nu_R\ar\nu_L}.$ Namely, it was argued in \cite{Kachelriess:2004ds} that the number of neutrinos from the neutronization burst is almost a model-independent value; the uncertainty in the neutrino event number at a Mt-scale detector was claimed to be (10-15)$\%$ for a supernova at 10 kpc. Therefore, if a considerable deficit of the neutronization burst neutrinos is revealed, it may be attributed to the $\nu_L\ar\nu_R$ transition of this neutrinos due to their precession in the interstellar magnetic field. This by itself may be considered as an evidence for the Dirac magnetic moment of neutrino, though not a completely convincing one. The deficit is determined by $\nu_L\ar\nu_R$ transition probability $P_{\nu_L\ar\nu_R},$ which is equal to
$P_{\nu_R\ar\nu_L}.$  Thus one can extract this probability from the value of the deficit.  If the high-energy neutrinos from the same explosion are detected, not only the evidence for the Dirac magnetic moment of neutrino would be compelling, but it would be possible to {\it determine} $\mu$ from this observation, using a more precise version of eq.(\ref{event rate}).

However, it is worth noting that in some cases high-energy neutrino events from the supernova explosion may be observed, but the deficit of the neutronization burst neutrinos may not be clearly established, and visa versa. For example, the first situation may take place if at the moment of a supernova explosion no Mt-scale detector is operational. In this case only a few neutronization burst neutrinos can be detected by Super-Kamiokande and other detectors of the similar sensitivity. An opposite possibility may be realised if a Mt-scale detector registers neutrinos from a supernova which possesses a strong magnetic field. In this case condition (\ref{Voloshin condition})  may be violated, and one may observe the deficit of the neutronization burst neutrinos without detecting high energy neutrinos.

\section{Discussion} 

First of all, let us discuss eq.(\ref{event rate}). It is obtained from
\be\label{event rate detailed}
N= \kappa \cdot \int dE \left[\frac{M_{\rm H_2O}}{m_{\rm H_2O} }~\frac{\sigma(E)}{4\pi D^2}  \int dt \int d^3r \frac{dn_{\nu_R}}{dEdt}(E,\rho(r,t), T(r,t))\right].  
\ee
Here\\
$\kappa$  is the fraction of $\nu_{eR}$ converted to $\nu_{eL}$ in the interstellar space on the way from the supernova to the Earth (see section 2.1),\\
$M_{\rm H_2O}$ and $m_{\rm H_2O}$ are the fiducial mass of water and the mass of the H$_2$O molecule, correspondingly,\\
$D$ is the distance from the supernova,\\
$\sigma(E)$ is the cross section of the reaction $\nu_e + \Ox \ar e^- + \F$ for the neutrino with energy $E,$\\
${dn_{\nu_R}}(E,\rho(r,t), T(r,t))/{dEdt}$ is the number of rigt neutrinos with energy $E$ emitted per unit energy interval in unite time from unite volume of supernova matter with density $\rho(r,t)$ and temperature $T(r,t).$ This later quantity is integrated over the volume of the the supernova core and over the time of explosion. Two nontrivial points are to calculate ${dn_{\nu_R}}/{dEdt}$ for given $E,T, \rho,$ and to perform the integration over $d^3 r dt.$

The first point mainly deals with the calculation of probabilities of the neutrino spin flips on electrons and protons (reactions (\ref{spin flip reactions})) in a dense hot media. In \cite{Barbieri:1988nh} the media effects were accounted for by introducing the effective mass in the photon propagator. More sophisticated considerations \cite{Kuznetsov:2007mp}\cite{Ayala:1998qz} give a somewhat larger result for ${dn_{\nu_R}}/{dEdt}.$ It is not completely clear how the spin flip on nuclei contributes during different stages of supernova explosion (for example, Barbieri and Mohapatra\cite{Barbieri:1988nh} consider only spin flips on protons (and electrons) after the bounce, while Notzold\cite{Notzold:1988kz} relies completely on the spin flips on large nuclei before the bounce). 

To proceed with the second point, one, in general, needs to know the evolution of the state of matter during the explosion, i.e. $\rho(r,t)$ and $T(r,t).$ In ref.\cite{Barbieri:1988nh} an oversimplified model was used: both temperature and density were considered to be constant within a sphere of volume $V$ (supernova core) for a time interval $t.$ The following numerical values were used:\\
$V=4\cdot 10^{18}$ cm$^3$ (which corresponds to the core radius 10 km),\\
t=(0.5--1) s,\\
$\rho=8\cdot 10^{14}$ g/cm$^3$,\\
$T=(30-60)$ MeV.\\
From the modern point of view (see, for example, \cite{Dessart:2005ck}) typical density and temperature are overestimated in \cite{Barbieri:1988nh}. This could lead to the overestimation of the of the $\nu_R$ production rate. We are not aware of any calculations of this quantity which accurately account for the evolution of the state of the supernova core. Such calculations could greatly diminish the uncertainties in the predicted number of high energy neutrino events.

We think that underestimation of the spin flip probability and overestimation of the supernova core temperature and density in \cite{Barbieri:1988nh} could cancel each other to some extent. Nevertheless, in view of this uncertainties we used the lower value for $N$ from the interval suggested in \cite{Barbieri:1988nh}. 

We would like to point out that the detailed calculations of neutrino spin flip rate with supernova dynamics included are of interest not only in the case of Dirac neutrinos, but in the case of Majorana neutrinos also. Majorana neutrinos may have transition magnetic moments, which can lead to $\nu_e\ar\bar{\nu}_{\mu,\tau}$ spin flips incide the core. Non-electron antineutrinos, $\bar{\nu}_{\mu}$ and $\bar{\nu}_{\tau},$ may be then partially converted to electron antineutrinos, 
$\bar{\nu}_{e},$ due to the flavor transformations. As in the first tens of milliseconds from the beginning of the collapse only electron {\it neutrinos} are expected to be produced\footnote{In particular, this is the case for the beginning of the neutronization burst.}
 (see, for example, \cite{Kachelriess:2004ds}\cite{Dessart:2005ck}),
detection of electron {\it antineutrinos} at this stage would be a smoking gun for the Majorana neutrino magnetic moment.\footnote{This effect does not depend on the supernova magnetic field strength, and therefore differs from the similar one based on the $\nu_e\ar\bar{\nu}_{\mu,\tau}$ transition in the presence of the strong supernova magnetic field (see \cite{Ahriche:2003wt}-\cite{Akhmedov:2003fu} and references therein).}
From estimates presented in ref.\cite{Kuznetsov:2007mp} one can get that for $\mu=10^{-12}\mu_B$ and for the oversimplified supernova core model, described above, with $T=30$ MeV,  energy output of $\sim 150$ MeV flipped  neutrinos  is of order of $10^{51}$erg/30 ms. In the case of Majorana neutrinos this gives $\sim 10^{50}$erg/30 ms in cooled ($\sim 10$ MeV) non-electron {\it antineutrinos}. This is only one order of magnitude less than the energy output of the neutronization burst (which duration is (20--30) ms). However one should carefully take into account supernova dynamics, as well as antineutrino diffusion from the central part of the core to the neutrino sphere, in order to justify (or refute) this naive estimate.

There is an ongoing activity for detecting the Diffuse Supernova Neutrino Background (DSNB), i.e. neutrinos emitted by all previously exploded supernovae in the universe (see \cite{Malek:2002ns} for the most stringent experimental limit and \cite{Ando:2004hc} for the theoretical review). A question may arise whether the high-energy supernova neutrinos form a detectable analogue of the DSNB. Unfortunately, simple estimates show that this is unlikely.  A rough estimate for the event rate in the detector, $dN/dt,$ may be obtained:
\be
\frac{dN}{dt}\sim 4\pi (10 {\rm kpc})^2 R_0N \frac{c}{H},
\ee
where $c$ is the speed of light, $H=70$ km/s Mpc$^{-1}$ is the Hubble constant, $R_0\sim 10^{-4}$ yr$^{-1}$ Mpc$^{-3}$ is the local supernova rate, and N is defined above (see eq.(\ref{event rate})). For $\mu=10^{-12}\mu_B$ one gets ${dN}/{dt}\sim 1$ yr$^{-1}$ in the MEMPHYS detector. It is much smaller
than the expected rate of atmospheric neutrino events with energies less than 200 MeV, which is of order of 1000 yr$^{-1}$\cite{Ashie:2005ik}.

\section{Summary}

Neutrino signal from a future nearby supernova is likely to allow to probe Dirac neutrino magnetic moment $\mu$ up to $\mu \gtrsim 10^{-13} \mu_B$ at Super-Kamiokande and $\mu \gtrsim 0.5\cdot 10^{-13} \mu_B$ at a future Mt-scale detector. A smoking gun for such values of $\mu$ is the detection of (60-200)MeV neutrinos. With the Mt-scale detector another clear signature of the Dirac magnetic moment may be observed: the deficit of the neutronization burst neutrinos. If both signatures are observed, one is able to extract the value of the magnetic moment from the observations. 
\newline
\newline
{\bf\Large Acknowledgments}\\
The author is grateful to S.I.Blinnikov and M.I.Vysotsky for valuable comments. The
work was financially supported by the Dynasty Foundation scholarship, RF
President grant NSh-4568.2008.2, RFBR grants 07-02-00830-a and RFBR-08-02-00494-a.

\end{document}